\makeatletter
\@ifundefined{@parse@version@dash}{%
\def\@parse@version#1{\@parse@version@0#1}
\def\@parse@version@#1/#2/#3#4#5\@nil{%
\@parse@version@dash#1-#2-#3#4\@nil}
\def\@parse@version@dash#1-#2-#3#4#5\@nil{%
  \if\relax#2\relax\else#1\fi#2#3#4 }
}{}
\makeatother

\documentclass[%
 reprint,
 amsmath,amssymb,
 aps,
]{revtex4-2}

\usepackage{graphicx}
\usepackage{dcolumn}
\usepackage{bm}

\usepackage{color}

\begin{document}

\title{First-Order Antiferromagnetic Transition  and Novel Gapless Excitation in a 4$f$ Zigzag Chain Compound YbCuS$_{2}$}%

\author{Fumiya Hori$^1$,  Katsuki Kinjo$^1$, Shunksaku Kitagawa$^1$, Kenji Ishida$^1$, 
Yudai Ohmagari$^2$, and Takahiro Onimaru$^2$}
\affiliation{%
$^1$\mbox{Department of Physics, Kyoto University, Kyoto 606-8502, Japan} \\
$^2$\mbox{Department of Quantum Matter, Graduate School of Advanced Science and Engineering, Hiroshima University,} \\ Higashihiroshima 739-8530, Japan
}%

\begin{abstract}
We report on the $^{63/65}$Cu--nuclear magnetic resonance and nuclear quadrupole resonance (NQR)  studies of trivalent Yb zigzag chain compound YbCuS$_2$.
Sharp NQR signals were observed in the paramagnetic (PM) state.
Below $T_{\rm O} \sim$ 0.95 K, the multi peaks induced by the internal magnetic fields arising from the antferromagnetic (AFM) ordered moments appear and coexist with the PM signal down to 0.85 K, evidencing the first-order AFM phase transition at $T_{\rm O}$.
In addition, the nuclear spin-lattice relaxation rate $1/T_1$ abruptly decreases below $T_{\rm O}$ and shows the $T$-linear behavior below 0.5 K.
The significant large $1/T_1T$ value ($1/T_1T$ = 14 s$^{-1}$ K$^{-1}$) strongly suggests the presence of the novel gapless spin excitation in low temperature region. 
\end{abstract}

\maketitle

Frustration effects in low-dimensional quantum spin systems lead to
the suppression of long-range order and non-trivial ground states~\cite{frustrated-magnets}.
A typical example of such a frustrated quantum spin system is the $S$ = 1/2 zigzag-chain with competition
between the nearest-neighbor and next-nearest neighbor exchange interaction~\cite{Majumdar, Haldane}.
In this system, a variety of non-trivial quantum phenomena such as spin dimerization, a 1/3-magnetization plateau, and vector chirality have been theoretically suggested~\cite{zigzag1, zigzag2, zigzag3}
and experimentally confirmed in (N$_2$H$_5$)CuCl$_3$, Rb$_2$Cu$_2$Mo$_3$O$_{12}$, and so on~\cite{N2H5CuCl3, Rb2Cu2Mo3O12, zigzagcom, zigzag4}.

Recently, frustrated spin systems formed by rare-earth elements have been intensively studied.
The strong spin-orbit coupling and crystalline electric field (CEF) associated with 4$f$ electrons of rare-earth ions such as Ce$^{3+}$ and Yb$^{3+}$
lead to highly anisotropic exchange interactions, providing a novel platform for the spin system.
Many rare-earth-based frustrated compounds with kagom\'{e}, triangular, or honeycomb lattices have been studied.
For example, YbMgGaO$_4$~\cite{YbMgGaO4-1, YbMgGaO4-2, YbMgGaO4-3}
and NaYbSe$_2$~\cite{NaYbSe2}
with a triangular structure
exhibit an unusual ground state due to the frustration.
However, few rare-earth based compounds with a zigzag-chain structure have been found thus far.

YbCuS$_2$ is one such Yb-based zigzag-chain system. 
Figure~1(a) shows the crystal structure of YbCuS$_2$, which has an orthorhombic structure with the space group $P2_1 2_12_1$ (No. 19, $D^4_2$),  where Yb atoms form the zigzag chains along the $a$-axis~\cite{YbCuS2structure}.
The electrical resistivity of YbCuS$_2$ has been reported to have semiconducting behavior with an activation energy of 0.08 - 0.28~eV~\cite{YbCuS2resistivity}.
From the Curie-Weiss behavior of the magnetic susceptibility $\chi (T)$, the effective magnetic moment $\mu_{\rm eff}$ was estimated to be 4.62 $\mu_{\rm B}$, which is close to the value of the free Yb$^{3+}$ ion (4.54~$\mu_{\rm B}$).
In general, the CEF level is important for understanding high-temperature and high-field magnetic response in rare-earth systems. However, the low-temperature properties are governed by only the CEF ground state which was reported to be an isolated Kramers doublet~\cite{Ohmagari1, Ohmagari2}.
The negative Weiss temperature 
$\theta_{\rm p} = -31.6$~K with considering the CEF effect
indicates the antiferromagnetic (AFM) interaction between the Yb magnetic moments.
The magnetic specific heat divided by the temperature $C_{\rm m}/T$ shows a sharp peak at $T_{\rm O} \sim$ 0.95~K, suggesting a first-order (FO) phase transition.
Since the transition temperature $T_{\rm O}$ is much lower than $\theta_{\rm p}$, YbCuS$_2$ is expected to be a frustrated spin system.
Below 20~K, the magnetic entropy decreased from $R$ln2 expected for the Kramers doublet ground state. Here, $R$ is the gas constant.
The value of the magnetic entropy was only 20\% of $R$ln2 at $T_{\rm O}$, suggesting that the phase transition is suppressed by the fluctuation effect.

In addition, the unusual magnetic field $H$ dependence of $T_{\rm O}$ was reported~\cite{Ohmagari2}.
$T_{\rm O}$ is independent of $H$ up to 4~T and has a maximum at approximately 7~T, and there are three anomalies in the $H$-swept AC susceptibility measured at low temperatures up to 18~T.
These results suggest that a nontrivial ground state is realized in YbCuS$_2$.
%

To investigate the physical properties, particularly the origin of the $T_{\rm O}$ transition,
from a microscopic point of view,
we performed $^{63/65}$Cu-nuclear magnetic resonance (NMR) and nuclear quadrupole resonance (NQR) measurements on polycrystalline samples of YbCuS$_2$.
Our NQR results indicate that the FO AFM transition occurs at $T_{\rm O}$. 
Moreover, the nuclear spin-lattice relaxation rate $1/T_1$ of $^{63}$Cu at zero field abruptly decreases below $T_{\rm O}$ and exhibits $T$-linear behavior below 0.5 K,
suggesting the presence of novel gapless fermionic excitations.

\begin{figure}[t]
\centering
\includegraphics[width=9cm]{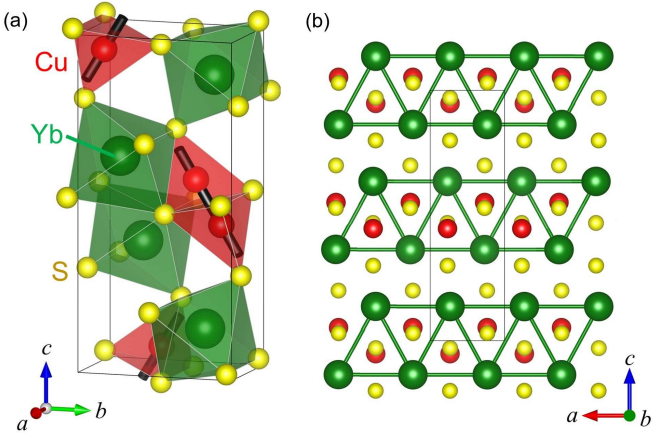}
\caption{(Color online) 
(a) Crystal structure of YbCuS$_2$ with the space group $P2_12_12_1$: the black sticks represent the principal axis of the electric field gradient tensor determined by calculation using the WIEN2k package.\cite{WIEN2k}
(b) View from the $b$-axis: zigzag chains along the $a$-axis are formed by Yb atoms.
The black box represents the unit cell.
The structural image was produced using the VESTA program~\cite{vesta}.}
\label{f1}
\end{figure}

Polycrystalline samples of YbCuS$_2$ were synthesized by the melt-growth method~\cite{Ohmagari1, Ohmagari2}.
The polycrystalline samples were coarsely powdered to increase the surface area 
for better thermal contact.
The powdered sample was mixed with GE 7031 varnish and solidified at zero magnetic field to avoid the preferential orientation of crystals in the NMR measurements and contact between crystals in the NQR measurements.    
A conventional spin-echo technique was used for the NMR and NQR measurements.
A $^3$He-$^4$He dilution refrigerator was used for the NQR measurements down to 0.075 K.
 $^{63/65}$Cu-NMR spectra
(with nuclear gyromagnetic ratios of $^{63}\gamma/2\pi$ = 11.289 MHz/T and $^{65}\gamma/2\pi$ = 12.093 MHz/T, respectively, and both with nuclear spin $I = 3/2$)
 were obtained as a function of $H$ at the frequency $f =$ 19.5 MHz. 
The principal axis of the electric field gradient (EFG) at the Cu site was determined by 
WIEN2k calculation using the density functional theory~\cite{WIEN2k}
since the principal axis was not determined experimentally due to a lack of a single crystal sample. The principal axis is represented by the black sticks in Fig.~1(a). 
$^{63/65}$Cu-$1/T_1$ was measured in YbCuS$_2$ and a reference compound LuCuS$_2$ to estimate the lattice contribution.
$1/T_1$ was evaluated by fitting the relaxation curve of the nuclear magnetization after the saturation to a theoretical function for the nuclear spin $I = 3/2$.
$1/T_1$ can be determined by a single relaxation component down to $T_{\rm O}$. 
However, the whole of the relaxation curve cannot be fitted to the single relaxation component below $T_\mathrm{O}$. Thus, we picked up the slowest components as shown in the supplemental material.

Figure~2(a) shows the $H$-swept $^{63/65}$Cu-NMR spectrum measured at 4.2~K on the powdered sample.
The quadrupole parameters $\nu_{zz}$ and $\eta$, which are explained subsequently, can be estimated by fitting the $H$-swept NMR spectrum to the simulated theoretical spectrum.
In general, the total effective NMR Hamiltonian of a nucleus in $H$ is given by
\begin{flalign}
\mathcal{H} =\mathcal{H}_{\mathrm{Z}}+\mathcal{H}_{\mathrm{Q}} 
&= -\frac{\gamma}{2 \pi} h (1+K) \boldsymbol{I} \cdot \boldsymbol{H} \nonumber \\
&+
\frac{h \nu_{z z}}{6}\left\{\left(3 I_{z}^{2}-I^{2}\right)+\frac{1}{2} \eta\left(I_{+}^{2}+I_{-}^{2}\right)\right\},
\end{flalign}
where $K$ is the Knight shift, $h$ is the Planck constant, and $I_{\pm}$ are the ladder operators of the nuclear spin $I$, which are defined as $I_{\pm} = I_x \pm iI_y$.
 $\nu_{zz}~(\propto V_{zz})$
is defined as $\nu_{zz} \equiv 3eV_{zz}Q/2I(2I-1)$ with the electric quadrupole moment $Q$. $\eta$ is an asymmetry parameter of the EFG defined as $\eta \equiv (V_{xx} -V_{yy} )/V_{zz}$, where $V_{ii}$ is the second derivative of the electric potential $V$ ($V_{ii} = \partial ^2 V/\partial x_{i}^2 $).
Since the NMR signal of each grain depends on the angle between the principal axis of the EFG in each grain and the direction of the magnetic field, the sum of the NMR signals for all solid angles could be observed in the non-oriented powder samples. 
The NMR spectrum was well fitted by the simulation with $K = 1.0$\%, $^{63}\nu_{zz} =$ 9.14~MHz, $^{65}\nu_{zz} =$ 8.48~MHz, and $\eta = 0.32$, as shown by the dashed line in Fig.~2(a).

We observed sharp $^{63/65}$Cu-NQR signals at $^{63}\nu_{\mathrm{Q}} = 9.28$~MHz and $^{65}\nu_{\mathrm{Q}} = 8.59$~MHz, as shown in Fig.~2(b).
The spectra were obtained by the frequency-swept method at 4.2~K without an external field. 
The obtained $^{63/65}$Cu-NQR frequencies $\nu_{\mathrm{Q}}$ at 4.2~K were consistently reproduced by 
$\nu_{\mathrm{Q}} = \nu_{zz}  \sqrt{1+\eta^2/3}$
with the quadrupole parameters obtained above.

\begin{figure}[t]
\centering
\includegraphics[width=8.5cm]{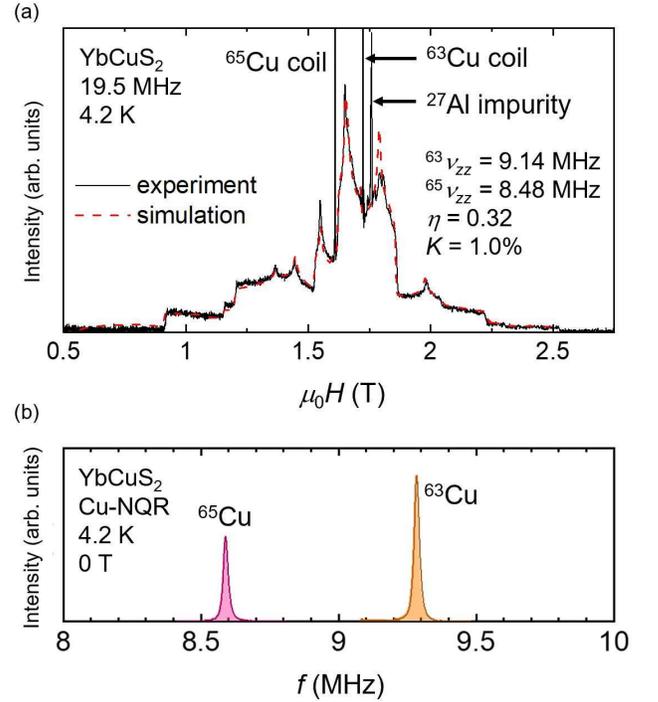}
\caption{(Color online)
(a) NMR spectrum obtained by the field-swept method at a fixed frequency of 19.5~MHz and 4.2~K: the black solid and (red) dashed curves are the experimental result and the simulation of the NMR spectrum, respectively (see text).
The three narrow lines in the NMR spectrum are signals from the Cu coil used for the NMR measurements and Al as an impurity arising from the NMR probe.
(b) $^{63/65}$Cu-NQR spectrum obtained by the frequency-swept method at 4.2~K.}
\label{f2}
\end{figure}

The occurrence of the FO AFM transition at $T_{\rm O}$ was concluded from the following NQR results. 
Figure~3(a) shows the frequency-swept $^{63/65}$Cu-NQR spectra measured at 4.2~K (upper panel) and 0.075~K (lower panel).
Each paramagnetic (PM) peak splits into 6 peaks at 0.075~K [(A1) - (A6) for $^{63}$Cu and (B1) - (B6) for $^{65}$Cu].
The $^{63}$Cu signals for (A1) and (A2) overlap with the $^{65}$Cu signals for (B5) and (B6), respectively.
As shown in Fig.~3(a), the observed $^{65}$Cu peaks almost coincides with the $^{63}$Cu signals scaled by the isotope ratio of the nuclear gyromagnetic ratio $^{65} \gamma /^{63} \gamma = 1.07$, not by that of the quadrupole moment $^{65}Q/^{63}Q = 0.93$.
In addition, since the first moments of the NQR spectrum does not change across $T_{\rm O}$, the EFG parameters are unchanged below $T_{\rm O}$. Therefore, the NQR signals are split due to the appearance of internal magnetic fields at the Cu site rather than to the change of electric factors such as charge density wave, charge-ordered, or structural transitions.

We performed spectrum simulations to estimate the magnitude of the internal magnetic fields and their orientation with respect to the principal axis of the EFG at 0.075~K.
As shown in the inset of Fig.~3(b), 
the 8 clearly visible peaks (A2) - (A5) and (B2) - (B5) 
can be fitted by the simulation with $\mu_0 H_\mathrm{int} = 0.021$ T, $\theta = 77$~deg, and  $\phi = 0$~deg,
where $\mu_0 H_\mathrm{int}$ is the absolute value of the internal fields, $\theta$  is the polar angle, and $\phi$ is the azimuthal angle of the internal fields from the principal axis of the EFG.
The small internal field at the Cu site suggests the tiny Yb ordered moment~\cite{small-Yb-ordered-moment}.
However, the peaks (A1), (A6), (B1) and (B6) 
cannot be fitted by above parameters: these peaks can be reproduced by assuming a slightly larger internal field.
Moreover, the presence of non-zero intensity between the peaks indicates the distribution of the internal fields.
The wide distribution of the internal fields suggests an incommensurate or spiral magnetic structure,
but a realized magnetic structure cannot be determined from the present measurement. Thus, elastic neutron scattering measurements are necessary, which will be performed in the near future.

The inset of Fig.~3(a) shows the temperature variations in the NQR spectra below 1.0~K.
Multi-peaks appear below $T_{\rm O} \sim$ 0.95~K and coexist with the PM peak of $\nu_{\mathrm{Q}} =$ 9.28~MHz.
The PM peak is not visible below 0.85~K. 
As shown in the main figure of Fig.~3(b), the internal field evaluated with the above simulation of the NQR spectra increases discontinuously below $T_{\rm O}$,
and the critical exponent $\beta$ derived by fitting the relation $H_\mathrm{int}(T) = H_\mathrm{int}(0)[(T_{\rm O} - T)/T_{\rm O}]^{\beta}$ is 0.05,
which is substantially smaller than the conventional mean-field value (0.5). 
These results indicate that the AFM transition is an FO phase transition, which is consistent with the sharp peak at $T_{\rm O}$ in the specific heat~\cite{Ohmagari2}.
The FO AFM transition is unusual and has been observed when the AFM transition and structural transition occur simultaneously~\cite{BaFe2As2, CaFe2As2}.
In YbCuS$_2$, since the NQR parameters are unchanged below $T_{\mathrm{O}}$, a structural transition is unlikely. 
Therefore, the FO AFM transition is likely to be related to the magnetic frustration.
In fact, a frustrated magnet EuPtSi shows a fluctuation-induced FO AFM transition without a structural phase transition~\cite{EuPtSi-1, EuPtSi-2}. The magnetic properties of YbCuS$_2$ deserves further investigation.

\begin{figure}[t]
\centering
\includegraphics[width=9cm]{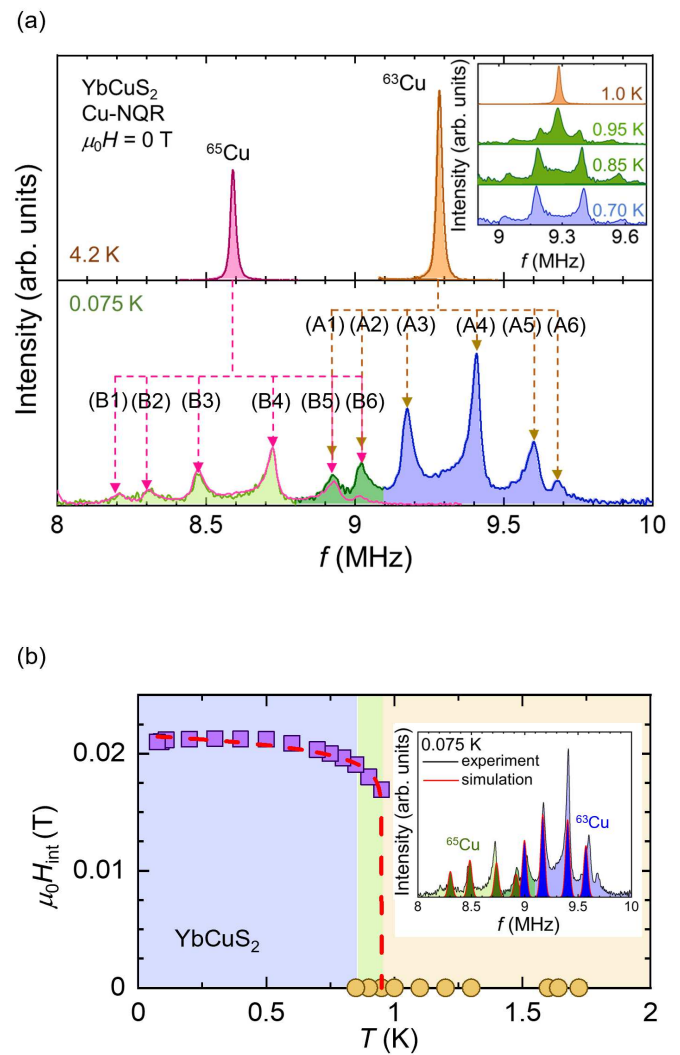}
\caption{(Color online) 
(a) $^{63/65}$Cu-NQR spectra obtained by the frequency-swept method at 4.2~K and 0.075~K: 
the pink curve represents the $^{63}$Cu signals scaled by $^{65} \gamma /^{63}  \gamma$ and shifted to the observed $^{65}$Cu signals, which is expected $^{65}$Cu signals with the appearance of internal fields.
The inset shows the temperature variations of the $^{63}$Cu-NQR spectrum for $0.7 \leq T \leq 1.0$~K.
(b) Temperature dependence of the internal fields $H_\mathrm{int}$ estimated by simulation of the NQR spectra:
the (red) dashed line represents the relation $H_\mathrm{int}(T) = H_\mathrm{int}(0)[(T_{\rm O} - T)/T_{\rm O}]^{\beta}$ with $\beta = 0.05$. 
The inset shows the frequency-swept $^{63/65}$Cu-NQR spectrum at 0.075~K, and 
the red solid curve is the simulation of the NQR spectrum.}
\label{f3}
\end{figure}

Figure~4 shows the temperature dependence of the nuclear spin-lattice relaxation rate $1/T_1$ measured at the $^{63}$Cu signal of YbCuS$_2$.
Below $T_{\rm O}$, $1/T_1$ was measured at the two peaks shown in the inset.
To investigate contributions other than the 4$f$ electrons, we also measured $1/T_1$ of a nonmagnetic reference compound LuCuS$_2$ and plot the results.
For LuCuS$_2$, the isotopic ratio of $1/T_1$ [$^{65}(1/T_1)/^{63}(1/T_1)$] is 0.88, which is close to the square of the quadrupole-moment ratio $(^{65}Q/^{63}Q)^2 \sim 0.86$. 
This indicates that in LuCuS$_2$, $1/T_1$ is determined by the electric quadrupole relaxation originating from the phonon dynamics.  The low-temperature $1/T_1$ of LuCuS$_2$ is quite small.
In contrast, for YbCuS$_2$, the isotopic ratio of $1/T_1$ $\sim 1.14$ coincides with the square of the gyromagnetic ratios $(^{65} \gamma/^{63} \gamma)^2 \sim 1.15$, indicating that $1/T_1$ is determined by the magnetic relaxation from the Yb$^{3+}$ moments.
$1/T_1$ exhibits a broad maximum at approximately 50 K, where the magnetic entropy reaches $R \ln 2$, as expected for the Kramers doublet ground state in specific-heat measurements.
Note that $1/T_1$ gradually decreases below 50 K, which may  be related to the entropy release observed in the specific-heat measurements~\cite{Ohmagari2}.
Below 5 K, $1/T_1$ becomes constant, suggesting that the local moments fluctuate with a short-range correlation. This behavior agrees with the theoretical prediction for the one-dimensional (1-D) spin chain~\cite{Sachdev}, and it has been observed in the $S$ =1/2 1-D cuprate antiferromagnet~\cite{Ca2CuO3, Sr2CuO3}. 
On the other hand, the absence of the critical slowing down behavior around $T_{\rm O}$ is quite a contrast to the divergence behavior of $1/T_1$ observed in the zigzag chain compound CaV$_2$O$_4$~\cite{CaV2O4}
and 1-D cuprate antiferromagnets near $T_{\rm N}$~\cite{Ca2CuO3, Sr2CuO3} 
where the magnetic order occurs due to interchain coupling.   
The absence of the slowing down behavior is likely to be related to the character of the FO phase transition~\cite{CaFe2As2}.

Below $T_{\rm O}$, $1/T_1$ decreases rapidly; 
it is roughly  proportional to
$T^5$
down to 0.5~K.
In general, $1/T_1$ of local-moment frustrated systems, such as the triangular Heisenberg antiferromagnet, is determined by the two-magnon process. It can be expressed as $1/T_1 \propto T^{2D-1}$ for $T \gg \Delta$, where $\Delta$ is the spin gap, and $D$ ($= 3$ or $2$) is the dimensionality of the spin-wave dispersion~\cite{Moriya1, Moriya2, Maegawa, NiGa2S4}.
In YbCuS$_2$,
$D = 3$ is likely.
Conversely, peculiar $T$-linear behavior was observed below 0.5~K.
Theoretically,
$1/T_1$ decreases exponentially [$1/T_1 \sim \exp{(-\Delta/T)}$] for  $T \ll \Delta$ at low temperatures in conventional semiconducting antiferromagnets with a spin gap.
Thus, the $T$-linear behavior suggests the presence of gapless excitation, which makes YbCuS$_2$ quite different from conventional antiferromagnets.

We note that similar $T$-linear behavior of $1/T_1$
below transition temperature was reported in several kagom\'{e} systems such as Zn-brochantite ZnCu$_3$(OH)$_6$SO$_4$~\cite{Zn-brochantite} and volborthite Cu$_3$V$_2$O$_7$(OH)$_2 \cdot $2H$_2$O~\cite{volborthite1, volborthite2}.
These compounds show the phase transition, and $T$-linear behavior of $1/T_1$ was observed below the transition temperature.
To explain these behaviors,
particle-hole excitations by spinons (fermionic elementary excitations), which are analogous to metallic excitations, were proposed~\cite{Zn-brochantite, volborthite1, volborthite2} since the experimental value of $1/T_1T$ is non-negligibly large.
In fact, we also point out that the experimental value of $1/T_1T$ in YbCuS$_2$ at 0.1~K ($\sim$ 14 s$^{-1}$ K$^{-1}$) is larger than that in a Cu-metal ($\sim$~0.83~s$^{-1}$ K$^{-1}$)~\cite{Cu-Korringa} by more than one order of magnitude.

Quite recently, novel gapless excitation has been observed in various semiconducting Kondo-lattice materials such as YbB$_{12}$~\cite{YbB12} and YbIr$_3$Si$_7$~\cite{YbIr3Si7}.
In these compounds, although the band gap opens at a low temperature due to the hybridization between the localized $f$ and the conduction electrons, quantum oscillation at high fields and the finite residual term in the specific heat and thermal conductivity experiments were observed~\cite{YbB12, YbIr3Si7}.
These results suggest the presence of gapless and charge-neutral excitations in the bulk properties, which are proposed to result in a quantum spin liquid with a spinon Fermi surface and the Majorana Fermi liquid.
We note the possibility that the large gapless excitation observed at low temperatures in YbCuS$_2$ might arise from such exotic Fermi liquid states, although YbCuS$_2$ is a conventional semiconductor.
To confirm this possibility, it is crucially important to prepare high quality single crystals for thermal conductivity and quantum oscillation measurements.        
%

\begin{figure}[t]
\centering
\includegraphics[width=9cm]{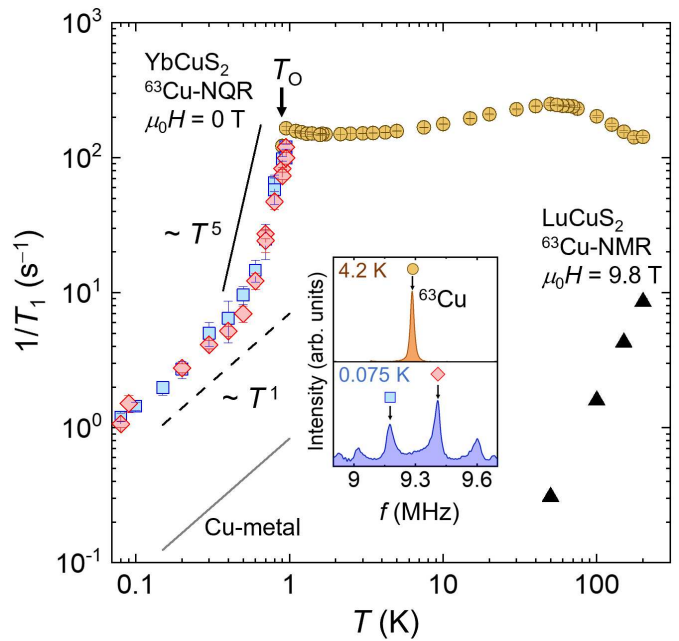}
\caption{(Color online) 
Temperature dependence of the nuclear spin-lattice relaxation rates $1/T_1$ of YbCuS$_2$ and a nonmagnetic reference compound  LuCuS$_2$:
the circles denote the $^{63}$Cu-NQR $1/T_1$ in the PM state of YbCuS$_2$.  
The diamonds and squares represent $1/T_1$ in the AFM state of YbCuS$_2$.
The black triangles indicate $^{63}$Cu-NMR $1/T_1$ in LuCuS$_2$ for $\mu_0 H =$ 9.8~T.
The gray solid line denotes $1/T_1$ in a Cu-metal~\cite{Cu-Korringa}.
The inset shows the $^{63}$Cu-NQR spectra measured at 4.2~K and 0.075~K.
The spectrum peaks at which $1/T_1$ was measured are denoted by the symbols.
}
\label{f3}
\end{figure}


In conclusion, we performed $^{63/65}$Cu-NMR/NQR measurements on powder samples of YbCuS$_2$ in which the Yb ions form a zigzag chain along the orthorhombic $a$-axis.
Below $T_{\rm O} \sim$ 0.95 K, multi peaks affected by the internal magnetic fields appear; they coexist with the PM signal down to 0.85~K, indicating that the FO AFM phase transition occurs at $T_{\rm O}$.
In addition, $1/T_1$ decreases abruptly below $T_{\rm O}$ and exhibits $T$-linear behavior below 0.5~K.
The significantly large $1/T_1T$ value--more than one order of magnitude larger than the value for metallic Cu--suggests the presence of novel gapless spin excitation originating from exotic fermions.
Our finding of the large gapless excitation unveils the presence of unknown fermionic quasiparticle in frustrated magnets.

\begin{acknowledgments}
The authors would like to thank Y. Maeno, S. Yonezawa, A. Ikeda, and Y. Matsuda for valuable discussions. 
This work was supported by the Kyoto University LTM Center, Grants-in-Aid for Scientific Research (Grant Nos. JP15H05745, JP17K14339, JP19K03726, JP16KK0106, JP19K14657, JP19H04696, JP19H00646, and JP20H00130) and Grant-in-Aid for JSPS Research Fellows (Grant No. JP20J11939) from JSPS.
\end{acknowledgments}

%


\begin{thebibliography}{39}
\bibitem [{sma()}]{small-Yb-ordered-moment}%
  \BibitemOpen
  \href@noop {} {}\bibinfo {note} {When we assumed that the coupling between
  the nuclear Cu spin and Yb moments arises from the classical dipolar
  interaction, the magnitude of the hyperfine interaction is estimated as
  0.07~T/$\mu_{\rm B}$, resulting in 0.3~$\mu_{\rm B}$/Yb. This suggests that
  the ordered moment is much smaller than the effective moment of the CEF
  ground state (1.3~$\mu_{\rm B}$/Yb)}\BibitemShut {NoStop}%
\end{thebibliography}
\end{document}